\documentclass[useAMS,usenatbib]{mn2e}

\usepackage{aas_macros}
\usepackage{graphicx}
\usepackage{amsmath}
\usepackage{amssymb}
\usepackage{xspace}
\usepackage{ifthen}
\usepackage{lscape}
\usepackage{placeins}
\usepackage{longtable}

\usepackage{array}
\newcolumntype{L}[1]{>{\raggedright\let\newline\\\arraybackslash\hspace{0pt}}p{#1}}
\newcolumntype{C}[1]{>{\centering\let\newline\\\arraybackslash\hspace{0pt}}p{#1}}
\newcolumntype{R}[1]{>{\raggedleft\let\newline\\\arraybackslash\hspace{0pt}}p{#1}}

{ %<- modify value to suit your needs

\title[Simultaneous multi-band transits of WD 1145+017]{Simultaneous infrared and optical observations of the transiting debris cloud around WD 1145+017}

\author[Zhou et al.]
{\parbox{\textwidth}
  {G. Zhou$^{1}$\thanks{E-mail: \texttt{george.zhou@cfa.harvard.edu}},
L.~Kedziora-Chudczer$^{2,3}$,
J.~Bailey$^{2,3}$,
J.P.~Marshall$^{2,3}$,
D.D.R.~Bayliss$^{4}$,
C.~Stockdale$^{5}$,
P.~Nelson$^{6}$,
T.G.~Tan$^{7}$,
J.E.~Rodriguez$^{8}$,
C.G.~Tinney$^{2,3}$,
D.~Dragomir$^{9,18,19}$,
K.~Colon$^{10,11}$,
A.~Shporer$^{12,13}$,
J.~Bento$^{14}$,
R.~Sefako$^{15}$,
K.~Horne$^{16}$, and
W.~Cochran$^{17}$
\vspace{0.4cm}}\\
\parbox{\textwidth}{
$^{1}${Harvard-Smithonian Center for Astrophysics, 60 Garden St., Cambridge, MA 02138, USA}\\
$^{2}${School of Physics, University of New South Wales, Sydney, NSW 2052, Australia}\\
$^{3}${Australian Centre for Astrobiology, University of New South Wales, Sydney, NSW 2052, Australia}\\
$^{4}${Observatoire Astronomique de l'Universit\'{e} de Gen\`{e}ve, 51 ch. des Maillettes, 1290 Versoix, Switzerland}\\
$^{5}${Hazelwood Observatory, Victoria, Australia}\\
$^{6}${Ellinbank Observatory, Victoria, Australia}\\
$^{7}${Perth Exoplanet Survey Telescope, Western Australia, Australia}\\
$^{8}${Department of Physics and Astronomy, Vanderbilt University, 6301 Stevenson Center, Nashville, TN 37235, USA}\\
$^{9}${The Department of Astronomy and Astrophysics, University of Chicago, 5640 S Ellis Ave, Chicago, IL 60637, USA}\\
$^{10}${NASA Ames Research Center, M/S 244-30, Moffett Field, CA 94035, USA}\\
$^{11}${Bay Area Environmental Research Institute, 625 2nd St. Ste 209 Petaluma, CA 94952, USA}\\
$^{12}${Jet Propulsion Laboratory, California Institute of Technology, 4800 Oak Grove Drive, Pasadena, CA 91109, USA}\\
$^{13}${Sagan Fellow}\\
$^{14}${Research School of Astronomy and Astrophysics, the Australian National University, Canberra, ACT 2611, Australia}\\
$^{15}${SAAO, P O Box 9, Observatory 7935, South Africa}\\
$^{16}${SUPA Physics and Astronomy, University of St. Andrews, KY16 9SS, Scotland, UK}\\
$^{17}${McDonald Observatory, The University of Texas, Austin, TX 78712, USA}\\
$^{18}${Massachusetts Institute of Technology, Cambridge, MA 02139 USA}\\
$^{17}${NASA Hubble Fellow}\\
}}

\begin{document}

\date{Submitted 2016-04-24}

\pagerange{\pageref{firstpage}--\pageref{lastpage}} \pubyear{2016}

\maketitle

\label{firstpage}

\begin{abstract}
We present multi-wavelength photometric monitoring of WD 1145+017, a white dwarf exhibiting periodic dimming events interpreted to be the transits of orbiting, disintegrating planetesimals. Our observations include the first set of near-infrared light curves for the object, obtained on multiple nights over the span of one month, and recorded multiple transit events with depths varying between $\sim 20$ to 50 per cent. Simultaneous near-infrared and optical observations of the deepest and longest duration transit event were obtained on two epochs with the Anglo-Australian Telescope and three optical facilities, over the wavelength range of 0.5 to $1.2\,\mu\mathrm{m}$. These observations revealed no measurable difference in transit depths for multiple photometric pass bands, allowing us to place a $2\sigma$ lower limit of $0.8\,\mu\mathrm{m}$ on the grain size in the putative transiting debris cloud. This conclusion is consistent with the spectral energy distribution of the system, which can be fit with an optically thin debris disc with minimum particle sizes of $10^{+5}_{-3}~\mu$m.
\end{abstract}


\section{Introduction}
\label{sec:introduction}

The transit events detected around WD 1145+017 may be the first direct evidence for in-falling planetesimals polluting the atmosphere of a white dwarf. Photometric monitoring of the white dwarf by the \emph{K2} mission \citep{2014PASP..126..398H} and subsequent ground-based follow-up revealed a series of asymmetric transits, with depths upto 40 per cent and periods of 4.5 to 4.9 hours \citep{2015Natur.526..546V}. The white dwarf also shows heavy element pollution in its spectrum, and infrared excess in its spectral energy distribution (SED) \citep{2015Natur.526..546V,2016ApJ...816L..22X}. In addition, \citet{2016ApJ...816L..22X} reported that asymmetric broadened absorption lines from heavy elements were superimposed on the sharp atmospheric metal lines, suggesting the presence of circumstellar gas with line broadening velocities close to 300$\,\mathrm{km\,s}^{-1}$ surrounding the host star. Subsequent ground-based photometric monitoring has revealed a series of quasi-periodic dimming events \citep{2015Natur.526..546V,2015arXiv151006434C,2016ApJ...818L...7G,2016MNRAS.tmp..406R,2016arXiv160308823A}. Multiple events are often seen within a given period with depths that vary by as much as 60 per cent, and with life-times as short as days. These observations have been interpreted as a series of transits by fragments of a debris cloud surrounding WD 1145+017, and are perhaps indicative of dust originating from evaporating planetesimals. Some 30-50 per cent of white dwarfs have spectra that exhibit pollution by heavy elements \citep[e.g.][]{2003ApJ...596..477Z,2010ApJ...722..725Z,2014A&A...566A..34K}, while 1-3 per cent of white dwarfs have a detectable infrared excess suggestive of debris discs \citep[e.g.][]{2007ApJS..171..206M,2009ApJ...694..805F,2011MNRAS.417.1210G,2011ApJS..197...38D,2015MNRAS.449..574R}. WD 1145+017 shows signatures of heavy metal pollution and infrared excess, and is the first white dwarf found to exhibit transit events originating from its circumstellar material, giving us an unique opportunity to study the dust properties before it is accreted onto the star. 

If a significant part of the transiting debris cloud is optically thin, then we should expect the depth and shape of the transits to be wavelength dependent. \citet{2015arXiv151006434C} reported a series of simultaneous $V$ and $R$ band light curves from May 2015, finding no depth differences in the transits. \citet{2016arXiv160308823A} have reported OSIRIS spectrophotometric observations from the 10.4\,m Gran Telescopio Canarias. These spanned $0.48-0.92\,\mu\mathrm{m}$, and detected a series transits ranging from 25 per cent to 40 per cent, but which displayed no change in transit depth across multiple wavelength bins. This lack of wavelength-depth dependence in the optical excludes the presence of particles smaller than $ 0.5\,\mu \mathrm{m}$ in any debris cloud. 

Despite the faintness of WD 1145+017 ($g=17.0$, $J=17.5$\,mag), the deep transits make it an excellent target for multi-wavelength follow-up observations. In this study, we extend the wavelength baseline to the near-infrared with $J$ band light curves obtained at the Anglo-Australian Telescope (AAT) over multiple nights spanning a baseline of one month. Two of these near-infrared transits were observed simultaneously in the optical in coordination with other facilities. This, as well as subsequent optical follow-up observations with the Las Cumbres Observatory Global Telescope (LCOGT) network, allowed us to monitor the depths of the transits in different bands and observe their evolving appearances.

\section{Observations and reduction}
\label{sec:observations}

Photometric imaging observations have been obtained with multiple facilities: $J$ band infrared observations with the AAT; optical observations obtained simultaneously with some of this infrared data using a suite of small telescopes in different parts of Australia; and follow-up $g'$ band observations with the LCOGT network. These observations are summarised in Table~\ref{tab:observations}, and plotted together in Appendix~\ref{sec:lc_all}. Their acquisition and analysis are described as follows:

\begin{table*}
\centering
\caption{\label{tab:observations}List of observations}
\begin{tabular}{lllrrr}
\hline\hline
Facility & UT Date & UT Time & Photometric band & Band pass $(\mu\mathrm{m})$ $^a$& Exposure time (s) \\
\hline
AAT+IRIS2 3.9\,m & 2016 Feb 16 & 13:13 -- 18:16 & $J$ & $1.24_{-0.07}^{+0.09}$ & 30 \\
AAT+IRIS2 3.9\,m & 2016 Feb 17 & 16:27 -- 18:37 & $J$ & $1.24_{-0.07}^{+0.09}$ & 30 \\
AAT+IRIS2 3.9\,m & 2016 Feb 19 & 16:27 -- 18:34 & $J$ & $1.24_{-0.07}^{+0.09}$ & 30 \\
AAT+IRIS2 3.9\,m & 2016 Feb 20 & 16:59 -- 18:30 & $J$ & $1.24_{-0.07}^{+0.09}$ & 30 \\
AAT+IRIS2 3.9\,m & 2016 Feb 21 & 15:57 -- 18:34 & $J$ & $1.24_{-0.07}^{+0.09}$ & 30 \\
Hazelwood 0.32\,m & 2016 Mar 19 & 10:36 -- 18:18 & Clear & $0.58_{-0.23}^{+0.19}$ & 300 \\
AAT+IRIS2 3.9\,m & 2016 Mar 19 & 11:15 -- 16:26 & $J$ & $1.24_{-0.07}^{+0.09}$ & 30 \\
Ellinbank 0.32\,m & 2016 Mar 19 & 12:25 -- 16:48 & Clear & $0.58_{-0.23}^{+0.19}$ & 180 \\
PEST 0.30\,m & 2016 Mar 19 & 14:53 -- 20:28 & $V$ & $0.52_{-0.05}^{+0.06}$ & 240 \\
Hazelwood 0.32\,m & 2016 Mar 20 & 09:59 -- 13:15 & Clear & $0.58_{-0.23}^{+0.19}$ & 300 \\
AAT+IRIS2 3.9\,m & 2016 Mar 20 & 11:00 -- 17:27 & $J$ & $1.24_{-0.07}^{+0.09}$ & 30 \\
Ellinbank 0.32\,m & 2016 Mar 20 & 11:28 -- 17:02 & $R$ & $0.64_{-0.07}^{+0.04}$ & 240 \\
LCOGT 1\,m Sutherland & 2016 Mar 25/26 & 18:36 -- 02:00 & $g'$ & $0.47_{-0.06}^{+0.07}$ & 150 \\
LCOGT 1\,m Cerro Tololo & 2016 Mar 26 & 00:11 -- 05:11 & $g'$ & $0.47_{-0.06}^{+0.07}$ & 150 \\
LCOGT 1\,m Sutherland & 2016 Mar 28/29 & 18:36 -- 01:46 & $g'$ & $0.47_{-0.06}^{+0.07}$ & 150 \\
\hline
\end{tabular}
\begin{flushleft}
    $^a$ The wavelength centroid of the pass bands, accounting for detector efficiency, filter throughput, stellar flux, and atmospheric transmission. The error bars give the full width at half maximum of the transmission curve.\\
\end{flushleft}
\end{table*}

\subsection{AAT $J$ band observations}
\label{sec:obs_aat}

\begin{figure}
    \centering  
    \includegraphics[width=8cm]{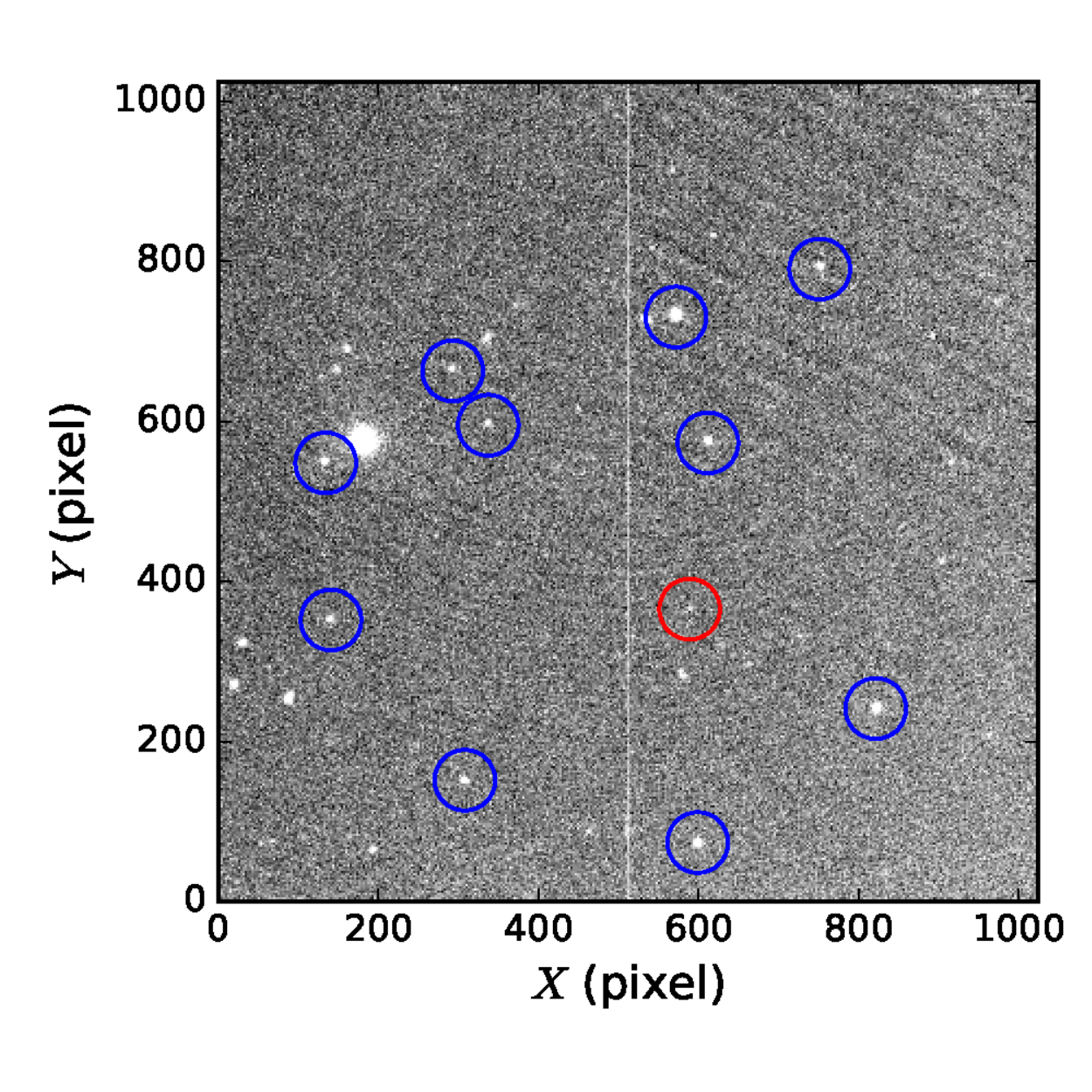}
    \caption{Example of a reduced image from the AAT+IRIS2 observations. The target star is marked by the red marker, the reference stars we used for relative photometry are marked by the blue markers. Note the size of ther marker does not represent the photometric aperture. The actual aperture was 5 pixels in radius, too small to be depicted on this diagram.}
    \label{fig:refstars_field}
\end{figure}

We obtained near-infrared light curves for WD 1145+017 using the IRIS2 camera \citep{2004SPIE.5492..998T} on the 3.9\,m AAT, located at Siding Spring Observatory, Australia. IRIS2 is a near-infrared camera, with a $1\,\mathrm{K} \times 1\,\mathrm{K}$ HAWAII-1 HgCdTe infrared detector, read out over four quadrants, achieving a field of view of $7\farcm7 \times 7\farcm7 $ and a pixel scale of $0\farcs4486\,\mathrm{pixel}^{-1}$. Our observations were performed in the $J$ band using a 30\,s integration time and read-out in double-read mode. The exposure times were set such that a photometric precision of at least $10\,\mathrm{per cent}$ was achieved at each exposure. The observing strategy, data reduction, and photometry extraction used followed the procedures described in \citet{2014MNRAS.445.2746Z,2015MNRAS.454.3002Z}. The one exception being that, due to the faintness of the target star, these observations were not defocused. Each sequence of light curves was obtained with the telescope locked to a nearby guide star, with the observer applying small manual offsets every $\sim 10$ minutes to ensure the target and reference stars stayed on the same pixel throughout. These target observations were bracketed by a series of dither sequences with defined offsets from the target. These were used to provide flat-field and sky background corrections. Photometry of the target and reference stars were extracted in circular apertures at a series of radii, and background estimated in annuli around each aperture. The image coordinate matching and photometric extraction are performed using the \emph{FITSH} package \citep{2012MNRAS.421.1825P}. Reference photometry is performed against a selection of stable reference stars. Six extraction apertures of fixed radii were used, with the aperture that yields the lowest out-of-transit scatter selected. An example IRIS2 image of the field, with the target and reference stars labelled, is presented in Figure~\ref{fig:refstars_field}. The transit depth analyses presented in Section~\ref{sec:lightcurve_model} were tested against all extraction apertures. The scatter in the derived transit depths due to aperture selection is smaller than the statistical error in each aperture by a factor of four or more. 

\begin{figure*}
    \centering
    \includegraphics[width=15cm]{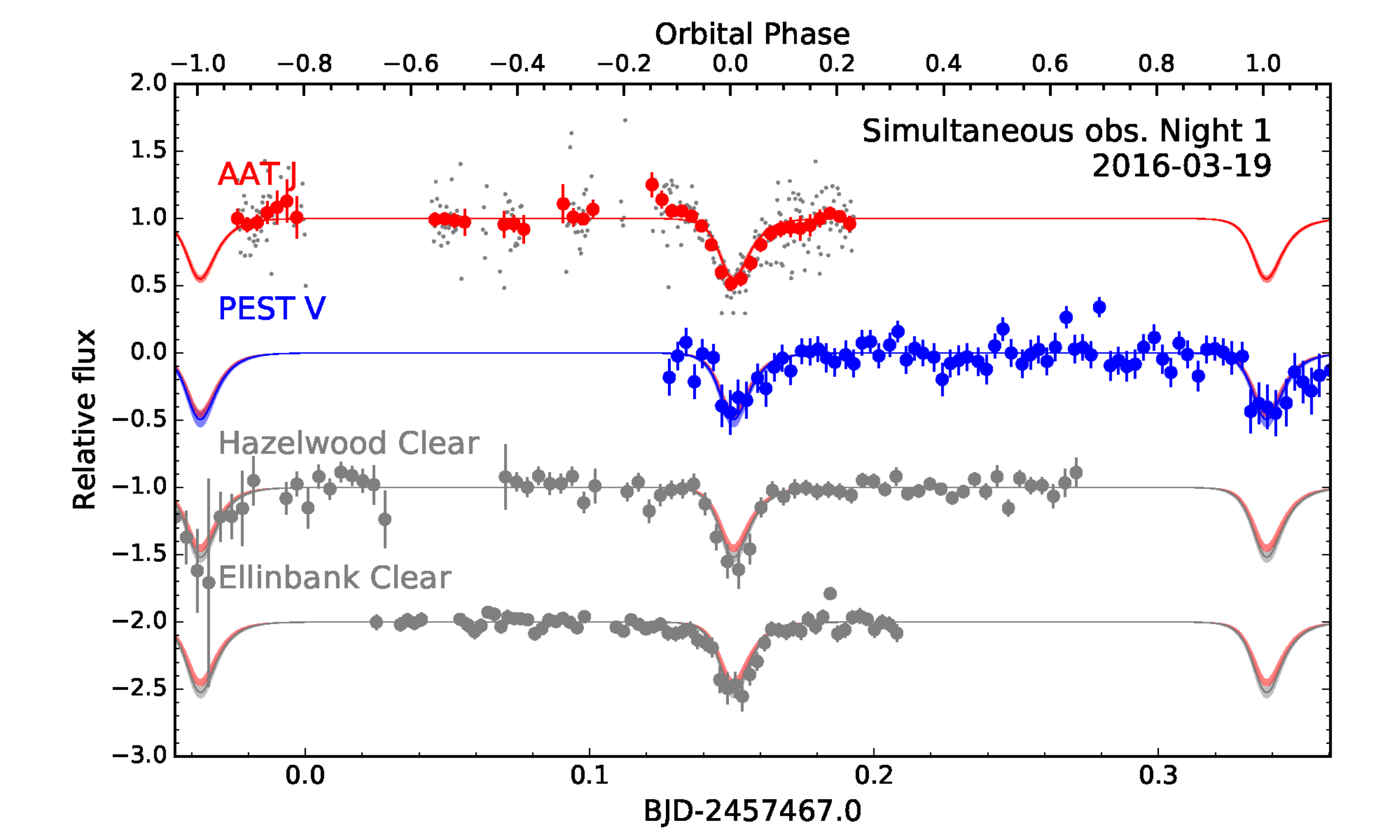}\\
    \includegraphics[width=15cm]{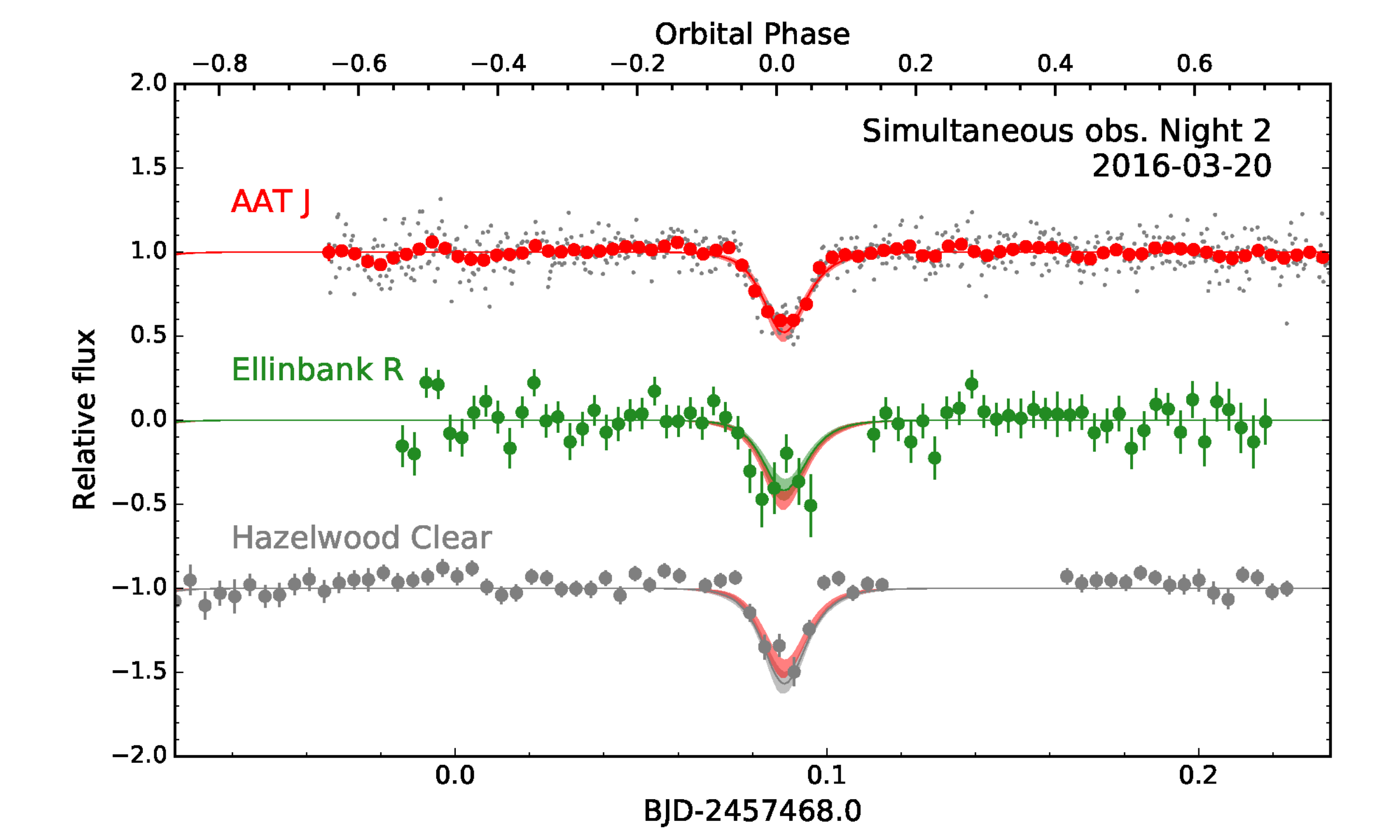}
    \caption{The light curves of WD 1145+017 from 2016 Mar 19 (top panel) and 2016 Mar 20 (bottom panel), simultaneously observed by the AAT in the infrared $J$ band, and by three small telescopes in the optical. The individual AAT observations are dotted in small points, and the 5 min binned light curve is plotted in red. The error bars represent the mean uncertainty of the points within the bin, scaled by the square root of the number of points per bin. The light curves of the optical facilities are shown at their native cadence. The fluxes from each set of observations are arbitrarily offset for clarity. The shaded areas represent the $1\sigma$ region in the model fit. The solid lines show the best fit models according to Equation~\ref{eq:model}. The optical light curves have the AAT $J$ band model fit (red) over-plotted for comparison. }
    \label{fig:lc_20160319}
\end{figure*}

The sets of AAT+IRIS2 observations are summarised in Table~\ref{tab:observations}. $J$-band observations were obtained on 2016 Mar 19 and 2016 Mar 20. These were taken simultaneous with optical observations from a suite of small telescopes (Section~\ref{sec:simultaneous-optical}). Poor weather meant only segments of light curves could be recovered on 2016 Mar 19. However conditions were photometric on 2016 Mar 20. These light curves are plotted in Figure~\ref{fig:lc_20160319}.

\begin{figure*}
    \centering
    \includegraphics[width=15cm]{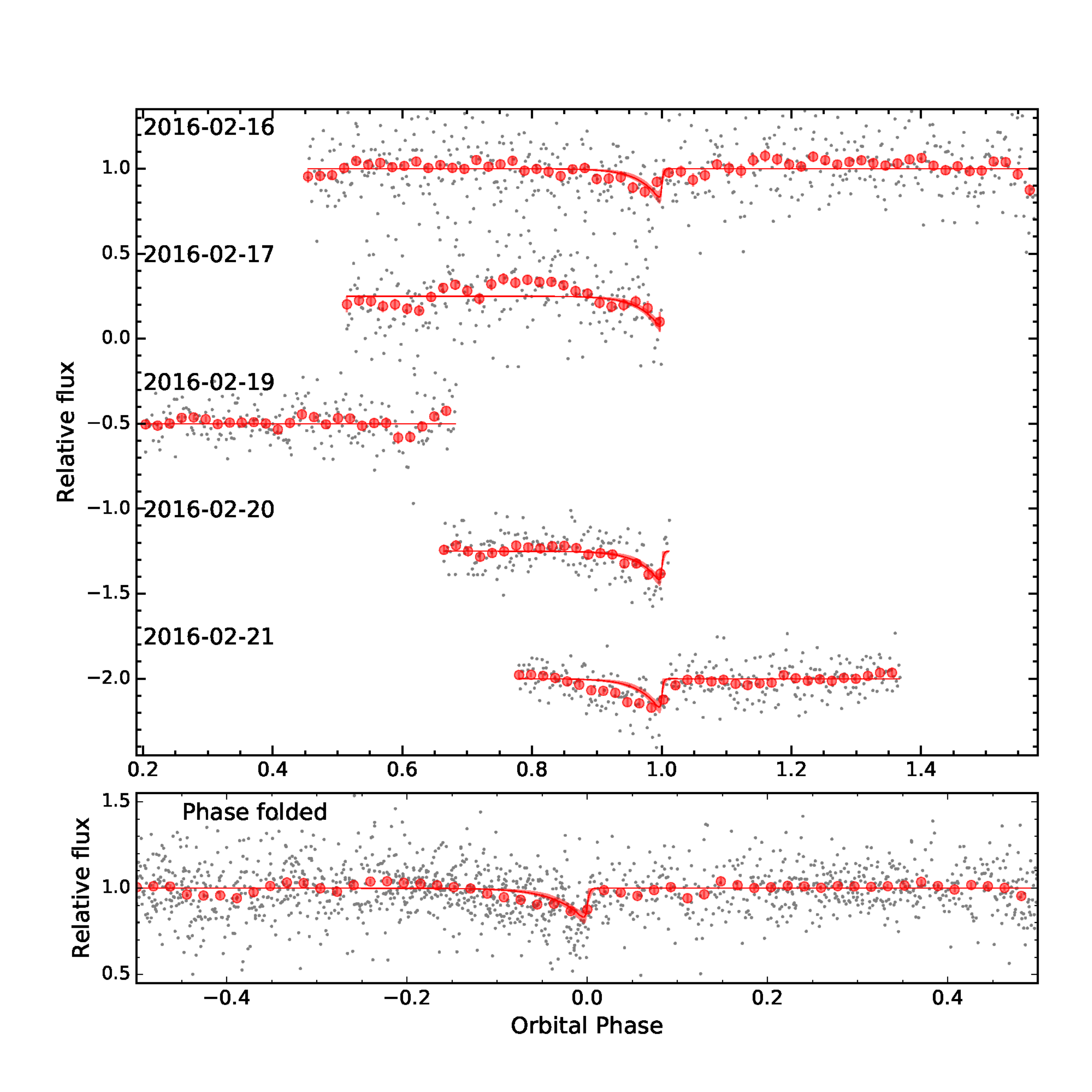}
    \caption{WD 1145+017 was monitored over five nights in 2016 Feb by the AAT in the near-infrared $J$ band only. These were analysed separately to the 2016 Mar simultaneous observations dataset. Two full transit events were captured during this period. The top panel shows the light curve from each night, with individual points plotted in grey, and the 5 min binned light curve in red. The shaded region represents the $1\sigma$ regime for the allowed models, assuming the transits can be described by a single model. The light curves from each night are arbitrarily offset for clarity. The bottom panel shows the phase binned light curve and model. }
    \label{fig:lc_201602}
\end{figure*}

Prior observations of WD 1145+017 were also obtained over five nights: 2016 Feb 16,17,19,20,21. These were not accompanied by simultaneous optical observations, and serve to provide a longer-term baseline to examine the near-infrared evolution of the debris cloud. These light curves are shown in Figure~\ref{fig:lc_201602}.

% \begin{figure*}
%     \centering
%     \caption{The light curves of WD 1145+017 from 2016-03-20, simultaneously observed in the infrared by AAT and in the optical by a series of small telescopes. See Figure~\ref{fig:lc_20160319} for description.}
%     \label{fig:lc_20160320}
% \end{figure*}

\subsection{Simultaneous optical observations}
\label{sec:simultaneous-optical}

Optical observations were obtained by multiple small telescopes across Australia on 2016 Mar 19 and 2016 Mar 20, simultaneously with the near-infrared photometry described in Section~\ref{sec:obs_aat}. 

Observations on these two nights were obtained with the 0.32\,m Planewave Corrected Dall-Kirkham (CDK) telescope at Hazelwood Observatory, operated by Chris Stockdale in Victoria, Australia. The exposures were made with a SBIG ST8XME $1.5\mathrm{K}\times1 \mathrm{K}$ CCD detector, yielding a $18'\times12'$ field of view and a $0\farcs73\,\mathrm{pixel}^{-1}$ plate scale. Due to the faintness of the target, no filter was used. The setup is fully described in \citet{2015arXiv150908953R}.

Observations on the same two nights were also obtained at the Ellinbank observatory, Victoria, with the 0.32\,m Planewave CDK telescope, operated by Peter Nelson. The observations were obtained using a SBIG 3200 ME CCD camera, with a $2184\times1472$ detector, with a field of view of $20\farcm2 \times 13\farcm5$ at a plate scale of $1\farcs12\,\mathrm{pixel}^{-1}$ when read out at $2\times 2$ pixel bin. The observations were performed over the optical (no filter) on 2016 Mar 19, and in the $R$ band on 2016 Mar 20. 

The Perth Exoplanet Survey Telescope (PEST) obtained observations on both nights. PEST is a fully automated observatory operating a 0.30\,m Meade LX200 Schmidt Cassegrain telescope, located in Perth, Western Australia and operated by T.G. Tan. The setup employs a SBIG ST-8XME detector, which provides a field of view of $31'\times 21'$ and a plate scale of $1\farcs2\,\mathrm{pixel}^{-1}$. Observations were performed in the $V$ band with an integration time of 240\,s on both nights. Unfortunately, the photometric precision from the 2016 Mar 20 observations were not sufficient to provide useful constraints on the $V$ band transit, and are not used in the subsequent analyses. The PEST facility is also described in \citet{2015arXiv150908953R}.

Each set of images was corrected for bias, dark current, and flat-fielded. Light curves were extracted from the reduced frames using aperture photometry via the \emph{FITSH} package as described in Section~\ref{sec:obs_aat}. Reference photometry was obtained using a set of reference stars that is largely consistent between facility to facility (though not fully consistent due to differences in field centre and field-of-view). The light curves are shown in Figure~\ref{fig:lc_20160319}.

\subsection{Subsequent optical follow-up from LCOGT}
\label{sec:lcogt}

\begin{figure*}
    \centering
    \includegraphics[width=15cm]{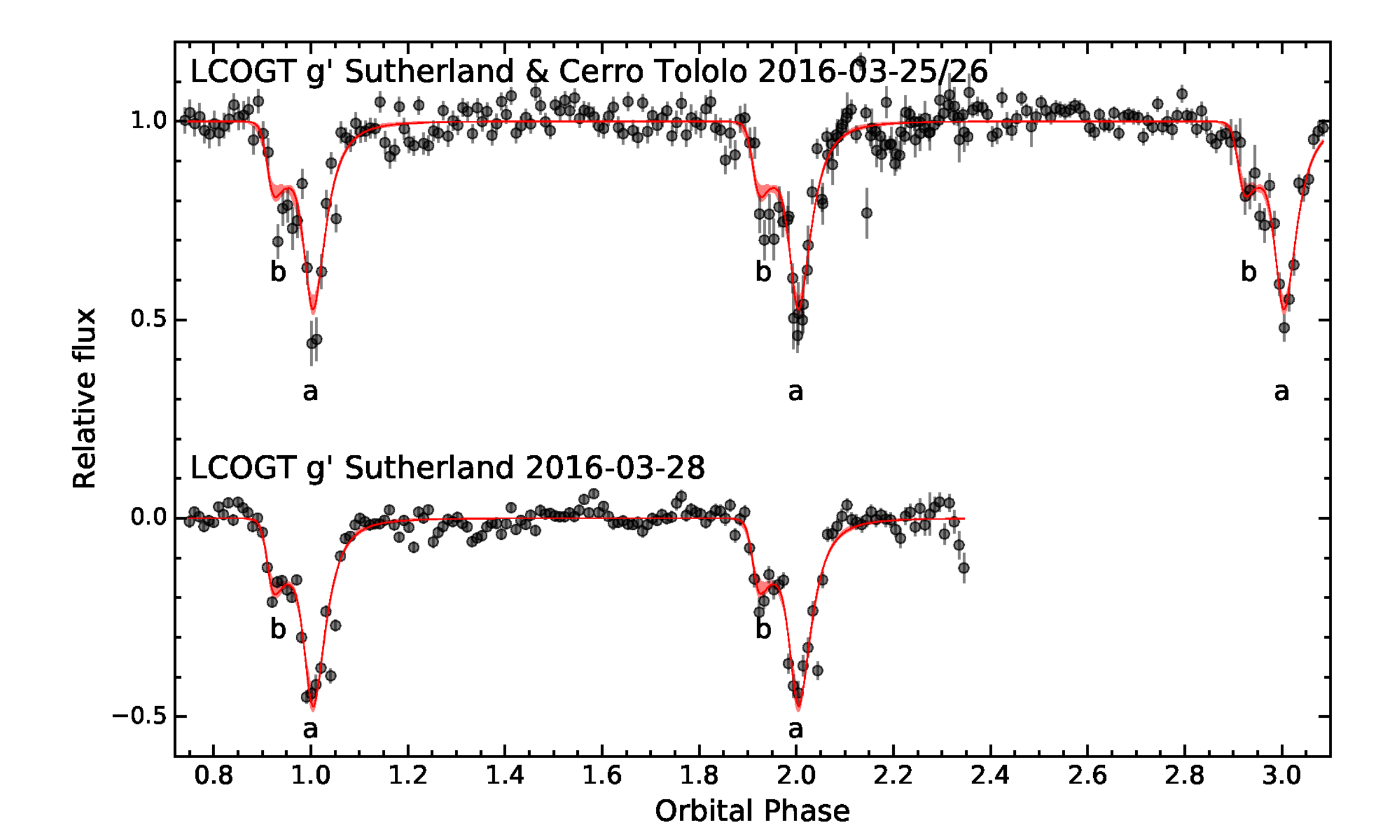}
    \caption{$g'$ band light curves of WD 1145+017 on 2016 Mar 25/26 and 2016 Mar 28, 5 days after the set of simultaneous observations. Five main transit events were observed over the two nights, each transit event involved a large 40 per cent event, and an adjacent 20 per cent event. Transits of the primary component $a$ and secondary component $b$ seen at each epoch are marked. The light curves were fit with a doublet transit model, allowing for $\tau_0$ and period to be shared. The light curves from each night are arbitrarily offset for clarity.}
    \label{fig:lcogt}
\end{figure*}

We obtained photometric observations using the LCOGT network \citep{2013PASP..125.1031B} within 5 days of our simultaneous infrared-optical campaign. The observations were performed on 2016 Mar 25 and 2016 Mar 28 using the 1\,m telescopes located at Sutherland observatory, South Africa, and on 2016 Mar 26 using an identical setup at Cerro Tololo observatory, Chile. An overlap of $\sim 2$ hours were available between the Cerro Tololo 2016 Mar 26 observations and that from Sutherland, allowing one transit to be captured simultaneously from both facilities. The observations were taken with the $4\mathrm{K} \times 4 \mathrm{K}$ SBIG STX-16803 camera, with a field of view of $15\farcm8\times 15\farcm8$ and a pixel scale of $0\farcs464\,\mathrm{pixel}^{-1}$ when read-out with $2\times 2$ pixel binning. The observations were performed in the SDSS $g'$ band to provide a large wavelength baseline for comparison. The raw frames were processed with the automated LCOGT pipeline. Photometric extraction and reference photometry were extracted according to Section~\ref{sec:obs_aat}, via the \emph{FITSH} package. The light curves are shown in Figure~\ref{fig:lcogt}.

\section{Light curve analyses}
\label{sec:lightcurve_model}

The analyses for the 2016 Mar 19 and 2016 Mar 20 simultaneous AAT near-infrared and optical observations are described in Section~\ref{sec:simultaneous_lc}, the derived parameters from this dataset can be found in Table~\ref{tab:simultaneous_params}. The additional sets of single-band $J$-band AAT light curves during five nights in 2016 Feb, and two nights of $g'$ band light curves with the LCOGT 1\,m network in March 2016 are analysed independently in Sections~\ref{sec:single_band_AAT} and \ref{sec:single_band_LCOGT}, respectively. The derived parameters from these observations are presented in Table~\ref{tab:single_obs}. 

\subsection{Modelling of simultaneous infrared and optical light curves}
\label{sec:simultaneous_lc}

The transits of WD 1145+017 evolve rapidly from orbit to orbit, and only simultaneous observations can give us a coherent analysis for the wavelength dependence of the transits. Following \citet{2014ApJ...784...40R} and \citet{2015arXiv151006434C}, we fit the transits with hyperbolic secant functions. Each transit is described by the parameters of reference transit centroid $(\tau_0)$, characteristic timescales for ingress $(\tau_1)$ and egress $(\tau_2)$, and scaling parameter $(C)$:
\begin{equation}
\label{eq:model}
  F(t) = 1 - C \left( \exp \left(\frac{-(t-\tau_0)}{\tau_1} \right) + \exp \left(\frac{t-\tau_0}{\tau_2} \right) \right)^{-1}\,.
\end{equation}
We assume a common shape for the light curve at different bands, such that the $\tau_0$, $\tau_1$ and $\tau_2$ parameters are shared across the transits observed on a given night. The depth parameter $C$ is left independent for each band, allowing us to probe for transit depth differences.

The best fit parameters and associated uncertainties are derived via a Markov chain Monte Carlo analysis, using the \emph{emcee} \citep{2013PASP..125..306F} affine invariant ensemble sampler. Since the transits are of short duration, we account for the long exposure time of the follow-up optical light curves by integrating over the models for each exposure.

The observations and model fits for 2016 Mar 19 and 2016 Mar 20 are shown in Figure~\ref{fig:lc_20160319}. The $1\sigma$ set of models allowed by the data are shaded for each plot. The resulting fit parameters are presented in Table~\ref{tab:simultaneous_params}. 

\begin{figure}
    \centering
    \includegraphics[width=7cm]{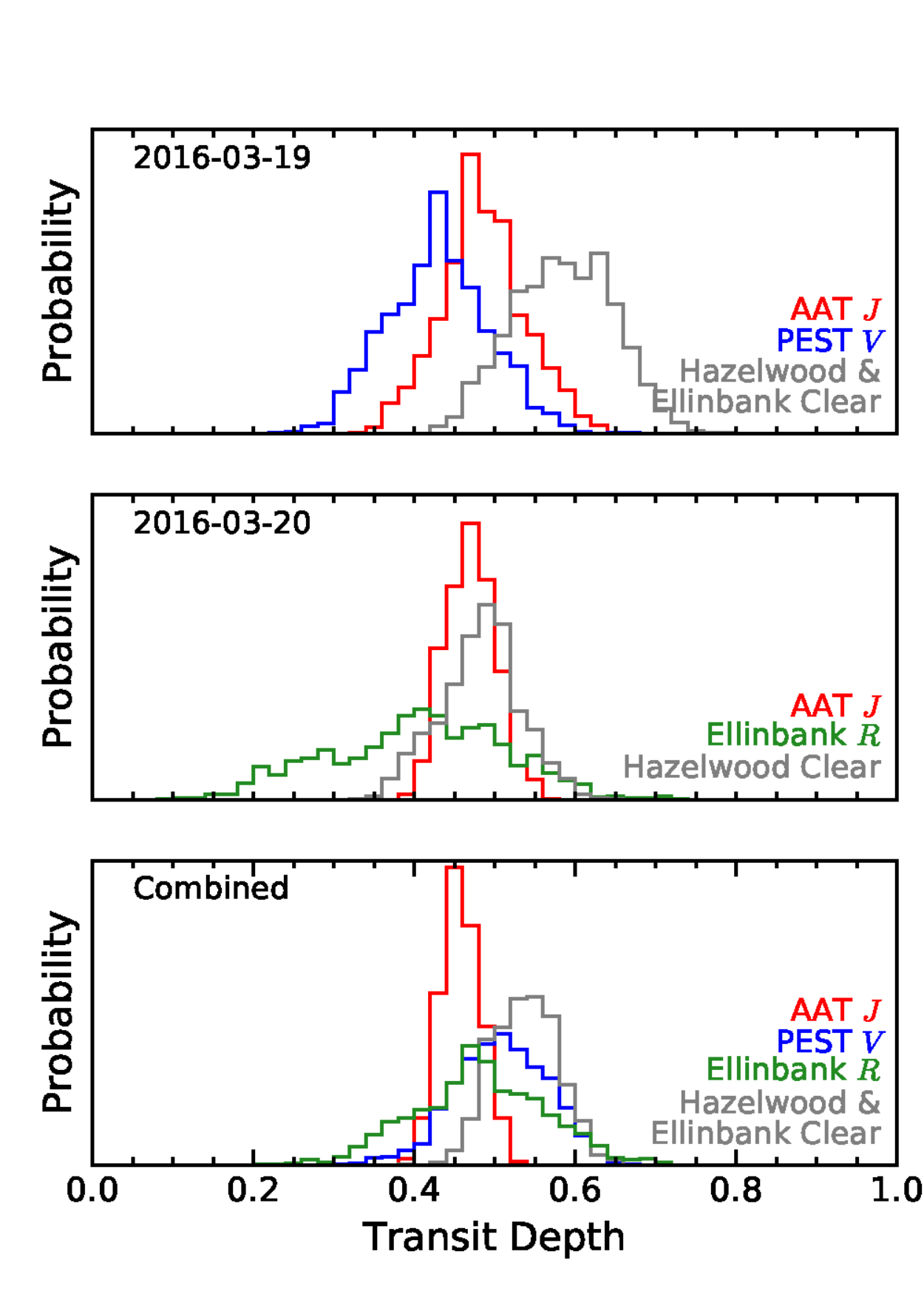}
    \caption{The transit depth posterior distribution for the transits simultaneously observed at the different photometric bands on 2016 Mar 19 (Top) and 2016 Mar 20 (Middle). The transits from each night were fit simultaneously, sharing the parameters governing the transit shape, but with different depths for each band. The transit depths agree between each photometric band, and between the two nights, at the $1\sigma$ level. The bottom panel shows the depth distribution derived from fitting the data from both nights simultaneously, with all transit shape parameters shared, and per-band transit depth independent. Again, the transit depths from each photometric band agree to within $1\sigma$.}
    \label{fig:depth_hist}
\end{figure}

In each case the transit depth determined at each pass band is consistent (given the uncertainties) with a single value. Following \citet{2015arXiv151006434C}, the depth $D$ (the minimum of the transit) is given by: 
\begin{equation}
D = C \frac{\zeta^{\zeta/(1+\zeta)} }{1+\zeta}\,,
\end{equation}
where $\zeta = \tau_2/\tau_1$. Figure~\ref{fig:depth_hist} shows the posterior distribution for the derived depths of the transits at each band for each of the nights. 

To check for consistency between different facilities, we also fitted the Hazelwood and Ellinbank 2016 Mar 19 clear band observations separately, deriving depths of $D_\mathrm{Clear} = 0.65_{-0.07}^{+0.08}$ and $0.58_{-0.05}^{+0.04}$, respectively. The two clear-band depths are self consistent to within $1\sigma$, but the Hazelwood depth differs from that of the PEST $V$ band depth by $2.3\,\sigma$ (even though the bandpass centroids are similar). Since the transits from the optical facilities are relatively low signal-to-noise, it is difficult to make interpretations from a single-facility, single band light curve. We therefore are prompted to combine all available light curves from the two nights into a joint analysis. 

A joint analysis of the two nights' data is only possible if there are no measurable transit shape changes between the nights. We find the best fit timescales to be $\tau_1 = 0.005\pm0.001$, $\tau_2 = 0.005\pm0.001$\,days on 2016 Mar 19 and $\tau_1 = 0.004\pm0.001$, $\tau_2 = 0.006\pm0.001$ on 2016 Mar 20. This is an important test, since the transit shape is known to change on a short time scale. The lack of detectable change allows us to combine the data in the same bands from the two nights.

We fit the data from both nights together, with all transits sharing the characteristic timescale parameters $\tau_1$ and $\tau_2$, and each band with an independent depth scaling parameter $C$. The reference transit time $\tau_0$ and transit period $P$ are also fitted for, and shared globally. The resulting transit depth posteriors are plotted in Figure~\ref{fig:depth_hist}. No significant difference in the transit depths were measured, with all depths consistent to within uncertainties. The full set of derived parameter values are given in Table~\ref{tab:simultaneous_params}. Interestingly, we also find no significant asymmetry for the transits, differing from the transits one month earlier (Section~\ref{sec:single_band_AAT}), and from many of the dimming events in literature \citep[e.g. FLWO transits from ][]{2015Natur.526..546V}. 

\begin{table*}
\centering
\caption{\label{tab:simultaneous_params}Transit parameters from multi-band simultaneous observations}
\begin{tabular}{lrrr}
\hline\hline
Parameter & 2016 Mar 19 & 2016 Mar 20 & Combined\\
\hline
\textbf{Fitted parameters} & & & \\
$\tau_0-2457400$ (BJD-TDB) & $67.150\,\left(_{-1}^{+1} \right)$ $^a$ & $68.086 \, \left(_{-2}^{+2} \right)$ & $67.1495 \, \left(_{-9}^{+10} \right)$\\
Period (d) &  &  & $0.1874\, \left(_{-1}^{+1} \right)$ \\ 
$\tau_1$ (d) & $0.0045\, \left(_{-9}^{+11} \right)$ & $0.0039 \,\left(_{-6}^{+7}\right)$ & $0.0037 \,\left(_{-4}^{+4} \right)$ \\
$\tau_2$ (d) & $0.0055\,\left(_{-7}^{+9}\right)$ & $0.006\,\left(_{-1}^{+1}\right)$ & $0.0056\,\left(_{-6}^{+7}\right)$ \\
$C_\mathrm{Clear}$ & $1.2_{-0.1}^{+0.1}$ & $0.9_{-0.1}^{+0.1}$ & $1.04_{-0.09}^{+0.08}$ \\
$C_V$ & $0.9_{-0.1}^{+0.1}$ &  & $1.0_{-0.1}^{+0.1}$\\
$C_R$ &  & $0.8_{-0.2}^{+0.2}$ & $0.9_{-0.1}^{+0.1}$ \\
$C_J$ & $0.95_{-0.09}^{+0.10}$ & $0.91_{-0.07}^{+0.05}$ & $0.88_{-0.04}^{+0.06}$ \\
\textbf{Derived depths} && \\
$D_\mathrm{Clear}$  & $0.59_{-0.06}^{+0.06}$ & $0.49_{-0.05}^{+0.04}$ & $0.54_{-0.04}^{+0.04}$ \\
$D_V$ & $0.43_{-0.06}^{+0.06}$  &  & $0.51_{-0.05}^{+0.05}$\\
$D_R$ &  & $0.41_{-0.12}^{+0.09}$ & $0.48_{-0.08}^{+0.08}$\\
$D_J$  & $0.48_{-0.04}^{+0.05}$ & $0.47_{-0.03}^{+0.03}$ & $0.46_{-0.02}^{+0.03}$\\
\hline
\end{tabular}
    \begin{flushleft}
    $^a$ Uncertainties in the parentheses are given for the last significant figure.\\
\end{flushleft}
\end{table*}

\subsection{Wavelength dependence}
\label{sec:wave_depend}

\begin{figure}
    \centering
    \includegraphics[width=7cm]{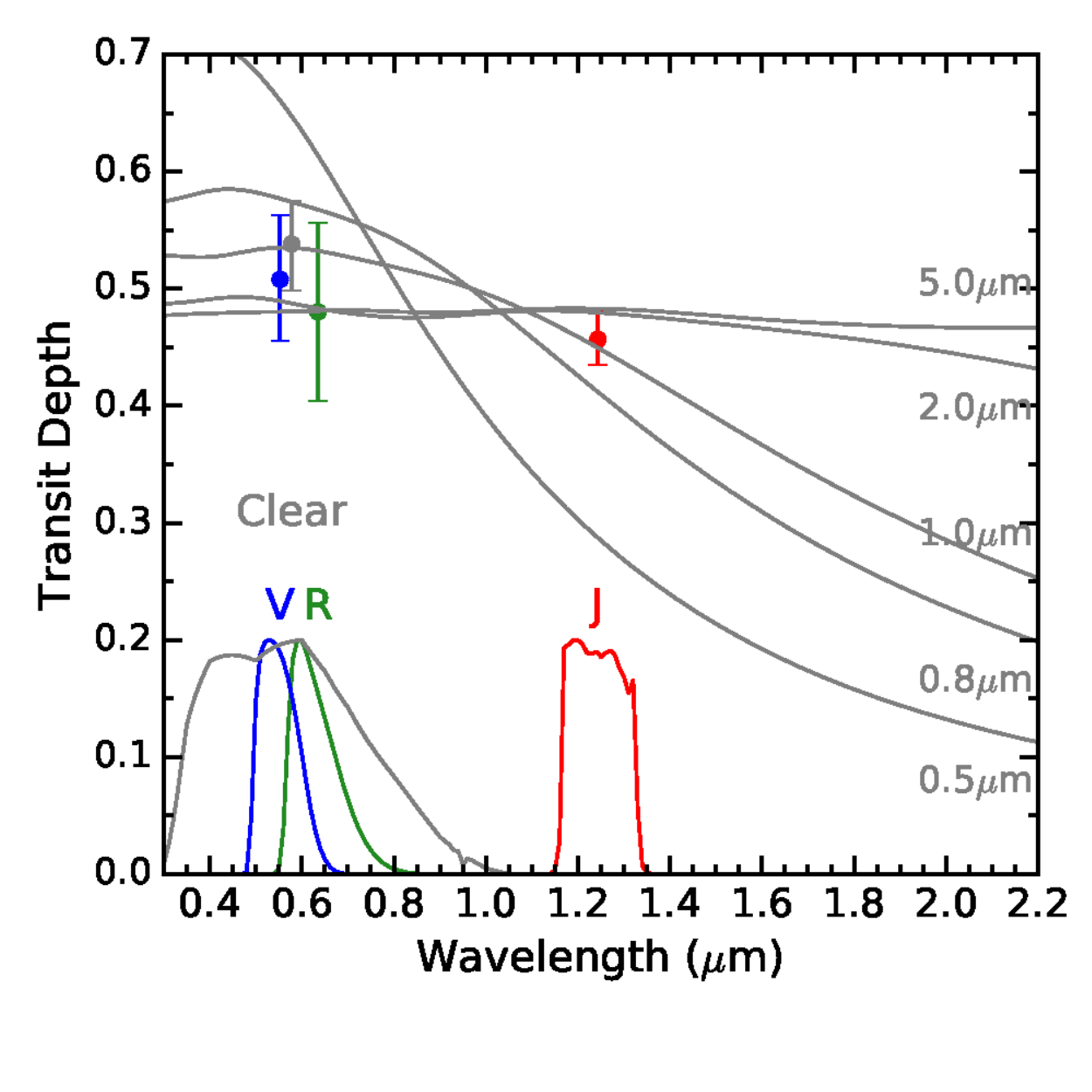}
    \caption{The transit depths of WD 1145+017 at the Clear, $V$, $R$, and $J$ photometric bands, as measured from simultaneous observations on 2016 Mar 19 and 2016 Mar 20. We find no strong evidence for any wavelength dependence of the transit depth. Extinction curves, calculated as described in Section~\ref{sec:wave_depend}, for particles 0.5 to 5.0\,$\mu$m are plotted for comparison. The particle sizes can be constrained to be $\gtrsim 0.8\,\mu$m.}
    \label{fig:band_depth}
\end{figure}

The transit depths, derived from the 2016 Mar 19 and 2016 Mar 20 transits are plotted in Figure~\ref{fig:band_depth}. The photometric band passes, scaled by flux of the white dwarf, are also plotted. If we assume a linear fit can describe the transit depths as a function of wavelength, then the depths derived from the simultaneous depth fits to the two nights are consistent with a blue-ward slope to the transit depths of $-0.11_{-0.04}^{+0.06}\,\mu\mathrm{m}^{-1}$, at $2\sigma$ significance. When each night was analysed independently, the $2\sigma$ signal of the blue-ward slope is only detectable in the 2016 Mar 19 dataset. The uncertainty on the slope was larger on the 2016 Mar 20 dataset, and no blue-ward trend was detected. In addition, given the variable nature of the transit depth and shape, and the low significance of the detection, this tentative blue-ward slope should be treated with caution.

To place constraints on the dust distribution in the debris cloud, we calculate a set of extinction cross sections for a power-law distribution of spherical particles \citep[as defined by ][]{1974SSRv...16..527H} with effective radii from 0.05 to 20.0\,$\mu$m, and effective variance of 0.1. The optical properties are those of `astronomical silicate' \citep[as per ][, at $\lambda = 0.5\,\mu$m these are $n = 1.694$ and $k = 0.02973$]{1984ApJ...285...89D,1993ApJ...402..441L}. The extinction was calculated using the Lorenz-Mie scattering code of \citet{2002sael.book.....M}, assuming the entire transiting cloud is optically thin. In the optical, a blue-ward slope is expected for distributions of small particles due to the Rayleigh scattering effect. No detectable wavelength dependence is expected of dust clouds consisting of larger particles in the spectral range we surveyed. With these assumptions, we calculate the $\chi^2$ between the set of extinction cross section curves and the observed transit depth -- wavelength variations. We rule out (with $2\sigma$ significance) particles smaller than $0.8\,\mu\mathrm{m}$; this is consistent, and slightly more constraining, with previous studies \citep{,2015arXiv151006434C,2016arXiv160308823A}. We also note the $1\,\mu\mathrm{m}$ model is slightly preferred over that of larger particle sizes, but this tentative interpretation is heavily dependent on the robustness of the blue-ward trend. Figure~\ref{fig:band_depth} shows a selection of the models from 0.5 to 5.0\,$\mu$m to illustrate the expected extinction curves. 

For comparison, short period, asymmetric, variable transits detected about three main sequence stars to date \citep{2012ApJ...752....1R,2014ApJ...784...40R,2015ApJ...812..112S} are interpreted as debris clouds surrounding disintegrating / evaporating terrestrial planets. Observations by \citet{2014ApJ...786..100C} did not detect any depth-colour dependence for KIC 12557548 b, inferring particles larger than 0.5\,$\mu$m. \citet{2015ApJ...800L..21B} tentatively claimed a deeper transit in the $u'$ and $g'$ bands compared to that in the $z'$ band, suggestive of dust particles of 0.25 to 1 $\mu$m in size. \citet{2013A&A...557A..72B} used the scattering properties of the KIC 12557548 b transits to constrain the particle size to between 0.1 to 1.0 $\mu$m. Gran Telescopio Canarias spectrophotometric observations of K2-22b measured a slightly deeper transit in the bluer wavelength bins in at least one of the transits observed, suggestive of particle sizes of 0.2 to 0.4 $\mu$m \citep{2015ApJ...812..112S}.  In this context, the lack of strong colour dependence of the WD 1145+017 transits is largely consistent with conclusions drawn from these systems.

\subsection{Analysis of single-band near-infrared AAT observations}
\label{sec:single_band_AAT}

Figure~\ref{fig:lc_201602} shows the AAT $J$ band light curves obtained from the 2016 Feb observations. Two full transit events were identified on 2016 Feb 16 and 2016 Feb 21 light curves. The light curves were fit with the same procedure described in Section~\ref{sec:simultaneous_lc}, with the transit shape and depth assumed to be constant over the five nights. The transits were significantly shallower than those observed in March 2016 (with depths of $D=0.17\pm0.03$) and exhibit significantly asymmetric ingress and egress timescales. In fact, the same hyperbolic secant function cannot fully model the 2016 Feb 16 and 2016 Feb 21 transits. Fitting the two transit separately, we find the depth and shape to evolve over the 5 nights. The transit on 2016 Feb 16 has a depth of $D = 0.12_{-0.05}^{+0.14}$, while the transit on 2016 Feb 21 was deeper and longer, with $D = 0.19\pm0.03$. The period derived from the transits is not significantly different from that in Table~\ref{tab:simultaneous_params} from the 2016 Mar observations. Given the scatter in the orbital period, it is impossible to determine if this transit is of the same object as that observed in 2016 Mar.

\subsection{Analysis of the $g'$ LCOGT observations}
\label{sec:single_band_LCOGT}

Observations were also performed with the LCOGT 1\,m network on 2016 Mar 25/26 and 2016 Mar 28, $\sim 5$ nights after the AAT observations (Figure~\ref{fig:lcogt}). It appears the system had evolved significantly over these five days, with LCOGT light curves showing five distinct transit events, each being a composite of two transiting bodies. We fit the light curves with a double transit, allowing for a shared $\tau_0$ and period. The primary component (labelled `a' in Figure~\ref{fig:lcogt}) transits with a depth of $D = 0.41_{-0.03}^{+0.02}$. The secondary component (labelled `b' in Figure~\ref{fig:lcogt}) has a shallower transit of $D = 0.19_{-0.02}^{+0.01}$. In both cases, the egress is longer than ingress, suggesting a trailing tail-like feature. The two transits are offset by $0.0164 \pm 0.0007$ days ($\sim 24$ minutes). The period derived from the dataset is largely consistent with the period of the system from the AAT and optical observations on 2016 Mar 19/20, and the transit centroid agrees to within uncertainties. The depth of the primary event is 20 per cent shallower than that observed by the AAT five nights earlier, again indicative of significant evolution in the system. 

Fitting the two nights separately, we find a slight evolution in the depth of the secondary component, with $D=0.24\pm0.02$ on 2016 Mar 25/26, and $0.19\pm0.01$ on 2016 Mar 28. The primary component displayed no significant change in transit depth, nor did we detect any significant change in the transit timing offset between the primary and the secondary.

From the LCOGT light curves, the secondary component transited $0.0164 \pm 0.0007$ days earlier than the primary. The AAT and optical observations on 2016 Mar 19/20 did not detect the secondary fragment, nor were any asymmetry detected in the transits of the primary fragment. Assuming that the secondary component either fragmented from the primary, or were occulted by the primary during the AAT observations, we estimate that it has an orbital period $\sim 1\,\mathrm{min/orbit}$ shorter than that of the primary fragment, consistent with the drift rates of smaller fragments found by \citet{2016MNRAS.tmp..406R}.

\begin{table*}
    \caption{Derived parameters for single-band observations}
    \label{tab:single_obs}
    \centering
    \begin{tabular}{lrrrrrr}
    \hline\hline
        Dataset & $\tau_0-2457400$ & Period (d) & $\tau_1$ (d) & $\tau_2$ (d) & $D$ & Offset (d) \\
    \hline
        AAT $J$ 2016 Feb 16--21 & $35.1592\,\left(_{-9}^{+5}\right)$ & $0.18716\,\left(_{-3}^{+3}\right)$ & $0.006\,\left(_{-1}^{+1}\right)$ & $0.0003\,\left(_{-2}^{+3}\right)$ & $0.17_{-0.03}^{+0.03}$ &  \\
        AAT $J$ 2016 Feb 16 & $35.155\,\left(_{-4}^{+3}\right)$ &  & $0.003\,\left(_{-2}^{+4}\right)$ & $0.001\,\left(_{-9}^{+20}\right)$ & $0.12_{-0.05}^{+0.14}$ & \\
        AAT $J$ 2016 Feb 20 & $40.2123\,\left(_{-7}^{+6}\right)$ &  & $0.009\,\left(_{-2}^{+2}\right)$ & $0.0010\,\left(_{-4}^{+8}\right)$ & $0.19_{-0.04}^{+0.04}$ & \\
        LCOGT $g'$ 2016 Mar 25--28 $^a$& $73.3330\,\left(_{-3}^{+3}\right)$ & $0.18727\,\left(_{-1}^{+1}\right)$ & $0.0027\,\left(_{-2}^{+3}\right)$ & $0.0051\,\left(_{-2}^{+3}\right)$ & $0.43_{-0.02}^{+0.01}$ & ...\\
        $^b$&&& $0.0013\,\left(_{-2}^{+3}\right)$ & $0.010\,\left(_{-2}^{+2}\right)$ & $0.200_{-0.008}^{+0.010}$ & $-0.0159_{-0.0005}^{+0.0006}$\\ 
        LCOGT $g'$ 2016 Mar 25,26 $^a$ & $73.3335\,\left(_{-4}^{+5}\right)$ & $0.0017\,\left(_{-2}^{+3}\right)$ & $0.0032\,\left(_{-4}^{+5}\right)$ & $0.44_{-0.02}^{+0.02}$ & \\
        $^b$ &&& $0.0021\,\left(_{-6}^{+11}\right)$ & $0.012\,\left(_{-3}^{+2}\right)$ & $0.24_{-0.02}^{+0.02}$ & $-0.014_{-0.001}^{+0.001}$\\ 
        LCOGT $g'$ 2016 Mar 28 $^a$& $76.3295\,\left(_{-4}^{+4}\right)$ & $0.0035\,\left(_{-2}^{+3}\right)$ & $0.0056\,\left(_{-2}^{+3}\right)$ & $0.44_{-0.01}^{+0.01}$ & \\
        $^b$ &&& $0.0012\,\left(_{-1}^{+2}\right)$ & $0.006\,\left(_{-1}^{+1}\right)$ & $0.19_{-0.01}^{+0.01}$ & $-0.0163_{-0.0005}^{+0.0005}$\\         
    \hline
    \end{tabular}
    \begin{flushleft}
    $^a$ Primary component\\
    $^b$ Secondary component
    \end{flushleft}
\end{table*}

% \begin{table*}
%     \caption{Derived parameters for single-band observations}
%     \label{tab:single_obs}
%     \centering
%     \begin{tabular}{lrrrrrr}
%     Parameter & AAT $J$ 2016-02-16--21 & AAT $J$ 2016-02-16 & AAT $J$ 2016-02-20 & LCOGT $g'$ 2016-03-25--28 & LCOGT $g'$ 2016-03-25,26 & LCOGT $g'$     2016-03-28\\
%     \hline\hline
%     Period & $2457435.1592 \,\left(_{-9}^{+5}\right)$ & $2457435.155\,\left(_{-4}^{+3}\right)$ & $2457440.2123\,\left(_{-7}^{+6}\right)$ & $2457473.3330\,\left(_{-3}^{+3}\right)$ &  $2457473.3335\,\left(_{-4}^{+5}\right)$ & $2457476.3295\,\left(_{-4}^{+4}\right)$ \\
%     \hline
%     \end{tabular}
% \end{table*}

\section{Spectral energy distribution}
\label{sec:SED}

\begin{figure}
    \centering
    \includegraphics[width=9cm]{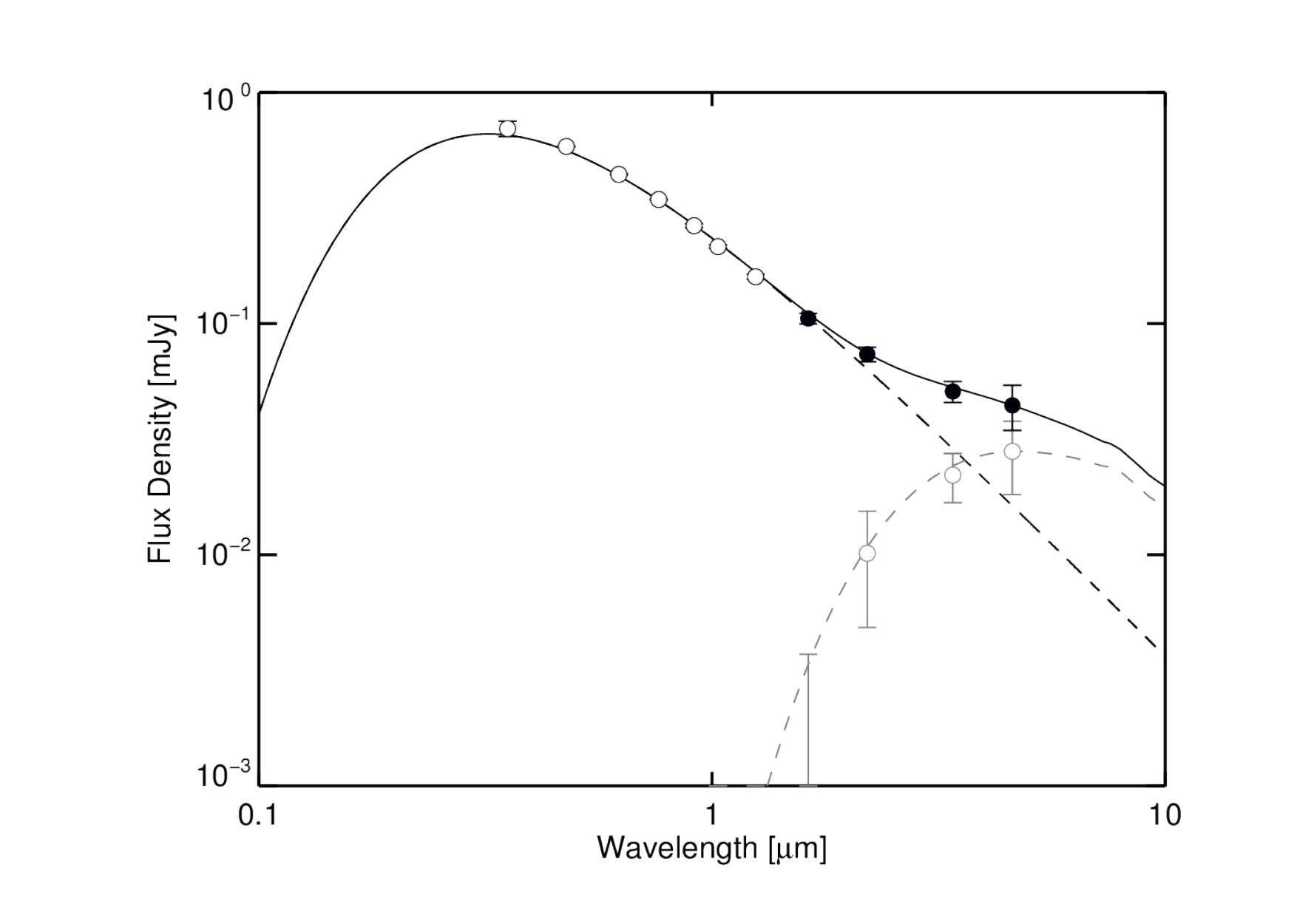}
    \caption{SED of WD 1145+017. White data points show non-excess measurements, whilst black data points denote wavebands with excess emission. Greyed out data points are the excess flux at the wavelength of observation. Uncertainties are 1-$\sigma$. The solid black line is the combined white dwarf+disc model. The dashed black and grey lines denote the white dwarf and disc contributions to the total. \label{fig:wd1145p017_sed}}
\end{figure}

Some 1-3 per cent of white dwarfs exhibit infrared excess, interpreted to originate from close-in debris disks surrounding the stars \citep[e.g.][]{2007ApJS..171..206M,2009ApJ...694..805F,2011MNRAS.417.1210G,2011ApJS..197...38D,2015MNRAS.449..574R}. WD 1145+017 exhibits significant infrared excess (Figure~\ref{fig:wd1145p017_sed}), thus providing an unique chance to examine the circumstellar material around an white dwarf via both transits and spectral energy distribution (SED) modelling. 

We use a list of photometric magnitudes from \citet{2015Natur.526..546V}, spanning optical and near-infrared wavelengths from 0.3 to 4.6~$\mu$m, combining measurements from SDSS \citep[$ugriz$;][]{2011ApJS..193...29A}, UKIDSS \citep[$YJHK$;][]{2007MNRAS.379.1599L}, and \textit{WISE} \citep[W1 and W2 -- W3 and W4 upper limits do not constrain the SED;][]{2010AJ....140.1868W}.% A summary of the photometry used in the modelling is presented in Table \ref{tab:wd1145p017_phot}. 

% \begin{table}
%     \centering
%     \caption{Photometry used in SED modelling of WD 1145+017. \label{tab:wd1145p017_phot}}
%     \begin{tabular}{ccc}
%         \hline\hline
%         Wavelength & Flux  & Reference \\
%         $[\mu \rm m]$ & [$\mu$Jy] & \\ 
%         \hline
%         \phantom{0}0.354 & 697~$\pm$~54 & 1\\
%         \phantom{0}0.477 & 583~$\pm$~21 & 1\\
%         \phantom{0}0.623 & 441~$\pm$~20 & 1\\
%         \phantom{0}0.763 & 344~$\pm$~22 & 1\\
%         \phantom{0}0.913 & 265~$\pm$~56 & 1\\
%         \phantom{0}1.031 & 215~$\pm$~37 & 2\\
%         \phantom{0}1.248 & 159~$\pm$~\phantom{0}4 & 2\\
%         \phantom{0}1.631 & 105~$\pm$~\phantom{0}6 & 2\\
%         \phantom{0}2.201 & \phantom{0}74~$\pm$~\phantom{0}5 & 2\\
%         \phantom{0}3.4\phantom{00} & \phantom{0}51~$\pm$~\phantom{0}5 & 3\\
%         \phantom{0}4.6\phantom{00} & \phantom{0}44~$\pm$~10 & 3\\
%         12.0\phantom{00} & $<$~\phantom{0}463 & 3\\
%         22.0\phantom{00} & $<$~5078 & 3\\
%         \hline
%     \end{tabular}
%     \begin{flushleft}
%     1. \cite{2011ApJS..193...29A}; 2. \cite{2007MNRAS.379.1599L}; 3. \cite{2010AJ....140.1868W}.
%     \end{flushleft}
% \end{table}

To model the dust emission we assume the white dwarf has the physical properties adopted in \cite{2015Natur.526..546V}, i.e. $d = 174~$pc, $T_{\rm eff} = 15,900~$K, $R_{\rm WD} = 1.4~R_{\oplus}$, and $M_{\rm WD} = 0.6~M_{\odot}$. We find the inferred luminosity of $L_{\rm WD} = 9.5\times10^{-3}~L_{\odot}$ is too low for there to be a blowout radius (i.e. a grain size below which the radiation force dominates gravitational force) for dust around the white dwarf \citep{1979Icar...40....1B}. 

However, the minimum size of dust grains around WD 1145+017 can be constrained from the thermal emission if we make a few simplifying assumptions. We adopt a model of an annular disc with dust grains in a power-law size distribution with a pure astronomical silicate composition \citep{2003ARA&A..41..241D}. We assume the disc is associated with the disintegrating planetesimal, and therefore is seen edge-on and is optically thin. These assumptions are supported by the lack of reddening at optical wavelengths in the white dwarf SED, and the white dwarf not being completed occulted during the transit events.

The model has numerous parameters and we have a limited number of excess data points to fit, so a few simplifying assumptions are made. We assume the disk radius and widths are set by the scatter in the period of the transiting planetesimal(s) \citep[4.5--4.9 hrs,][]{2015Natur.526..546V}, such that the disc has an inner edge, $R_{\rm in}=5.42\times10^{-3}$~AU ($1.16\,R_\odot$), and an outer edge of $R_{\rm out} = 5.96\times10^{-3}~$AU ($1.28\,R_\odot$), giving the disk a relatively compact geometry with width of $\Delta R/R \sim 0.1$. The size distribution of dust grains is fixed between a minimum size of $a_{\rm min}$ and maximum size of 1~mm. The dust is assumed to be in a steady state collisional cascade, such that the exponent of the size distribution power-law is 3.5 \citep{1969JGR....74.2531D}. 

A least squares fit of this simple model with $a_{\rm min}$ as the only free parameter produces a best fit of $a_{\rm min} = 10^{+5}_{-3}~\mu$m, with a $\chi^{2}_{red}$ of 1.5 (three free parameters). The disc fractional luminosity ($L_{\rm dust}/L_{\rm WD}$) is calculated to be 3.2$\times10^{-3}$. The dust mass inferred from the model is 1.3$\times10^{19}$~g, about 1 per cent of the mass in fragments calculated by \cite{2015Natur.526..546V} and \citet{2016MNRAS.tmp..406R}, determined from dynamical constraints on the periods of the transiting fragments. The grain size estimated here is consistent with that inferred for the dust from the flat transit depth between optical and near-infrared wavelengths. The absence of small grains, despite the lack of a blow-out radius, could be due to their rapid collisional destruction in the disc. Micron sized silicate grains have been inferred from the infrared excess around G29-38 \citep[e.g.][]{2005ApJ...635L.161R,2009ApJ...693..697R,2009AJ....137.3191J}, potentially due to tidal disruption of asteroidal material forming an debris disk.

Infrared excesses around other white dwarfs are often modelled via optically thick discs \citep[e.g.][]{2003ApJ...584L..91J,2009ApJ...694..805F}. In the case of WD 1145+017, an optically thick disc that exhibits strong infrared excess needs to be flared, warped, or misaligned with respect to the line of sight \citep{2015Natur.526..546V,2016ApJ...816L..22X}. However, if we assume that the disc is actually edge-on, then the fact that transits are observed in $J$ despite the significant flux from the disc implies it is optically thin. The lack of evidence for significant reddening in the optical also lends support to an optically thin disc interpretation, if the disk is aligned. In addition, the $300\,\mathrm{km\,s}^{-1}$ rotational broadening for the gas measured by \citet{2016ApJ...816L..22X} around the white dwarf would be co-spatial with the assumed location of our dust in the SED modelling, and the gas might reasonably be assumed to have been liberated from disintegrating planetesimals. The infrared excess of G29-38 has also been modelled with partially optically thin envelopes and disks \citep[e.g. G29-38,][]{2005ApJ...635L.161R,2009ApJ...693..697R}. The disk width and radius we assume for WD 1145+017 are slightly larger than previous white dwarf SED analyses. For example, \citet{2003ApJ...584L..91J} modelled the disc of G29-38 with an optically thick disc, finding $R_\mathrm{in} = 0.14\,R_\odot$ and $R_\mathrm{out} = 0.4-0.9\,R_\odot$, while \citet{2009ApJ...694..805F} and \citet{2014ApJ...786...77B} obtained disc radii of $\sim 0.3\,R_\odot$ for a series of white dwarfs that exhibited infrared excess. The blackbody radius for dust in the WD 1145+017 system is $\sim 0.9 R_\odot$, while real dust will necessarily be warmer than blackbodies at a given radius, and therefore bias the interpretation towards a closer-in disc. In addition, self-shadowing by the disc (from a flat geometry) or in dust clumps (where the transit depth implies a local $\tau \sim 0.4$) could produce a system where the dust appears cooler than expected given its radial location, allowing the disc to be closer in. We also note that a degeneracy exists between the inner radius and grain size of the disc; moving the inner radius closer towards the star will result in a less massive disc with larger minimum grain size. Future measurements of the mid-infrared and sub-millimetre flux of the system will help constrain the disc mass and dust properties of the system. The primary assumption of our model is that the disc originates from the disintegrating planetesimals at the $\sim 4.5$\,hr transit period. In the scenario where the current body is stirring up the remnants of a previous body, any relationship between the location and geometry of the dust disc and the transiting planetesimals around WD 1145+017 would be lost.

The disk of WD 1145+017 may be similar to other ‘hot dust’ stars \citep[e.g.][]{2013A&A...555A.104A,2014A&A...561A.114E,2016MNRAS.459.2893M}: main sequence stars with near-infrared excess emission, which have similarly hot dust in steady configurations \citep{2015Msngr.159...24E}. In these `hot dust' systems, the dust grains are very small $(a < 0.5 \, \mu\mathrm{m})$, such that they are inefficient absorbers of incident (optical) radiation which would keep them cooler than the dust temperature of $\sim 1100$\,K \citep{2016ApJ...816...50R}. In the case of hot dust stars, the small grains are trapped by the stellar magnetic field, preventing them from being blown outward. In the case of WD1145+017, there is no removal via radiation pressure/stellar wind force, so they may also persist for long timescales. 

\section{Conclusions}
\label{sec:conclusions}

We present multi-band photometric transits of debris clouds surrounding the white dwarf WD 1145+017. Simultaneous observations were obtained in the Clear, $V$, $R$, and $J$ bands on 2016 Mar 19 and 2016 Mar 20, recording two 50 per cent deep transit events. The depths derived from each band agreed to within $1\sigma$ on each individual night, and from the combined light curve of the two nights. The lack of transit depth -- colour dependence points to particulate sizes of $>0.8\,\mu\mathrm{m}$ (at $2\sigma$) in the transiting debris cloud.

WD 1145+017 also exhibits a significant infrared excess in its SED \citep{2015Natur.526..546V,2016ApJ...816L..22X}, allowing us to compare the inferred dust properties from the transit with that from the SED. We develop a simple model, describing the infrared excess with an optically thin disk in-line with the plane of the transiting debris clouds. The minimum particulate size derived from the SED modelling is $a_{\rm min} = 10^{+5}_{-3}~\mu$m, consistent with that found from the lack of colour-dependence in the transit depths.

Single-band photometric observations in the near infrared and optical, with the AAT and the LCOGT network, showed the evolution of the transit events over a one month baseline (See Appendix~\ref{sec:lc_all} for the ensemble set of of light curves). The transits were significantly shallower in 2016 Feb ($\sim 20$ per cent), and a secondary transiting fragment developed within the five days between the follow-up LCOGT observations and those from the AAT.  These observations demonstrated the fast changing nature of the transit events previously seen in \citet{2016MNRAS.tmp..406R}, \citet{2015arXiv151006434C}, and \citet{2016ApJ...818L...7G}.

\appendix
\section{Ensemble of light curves}
\label{sec:lc_all}

\begin{figure*}
    \centering
    \includegraphics[width=15cm]{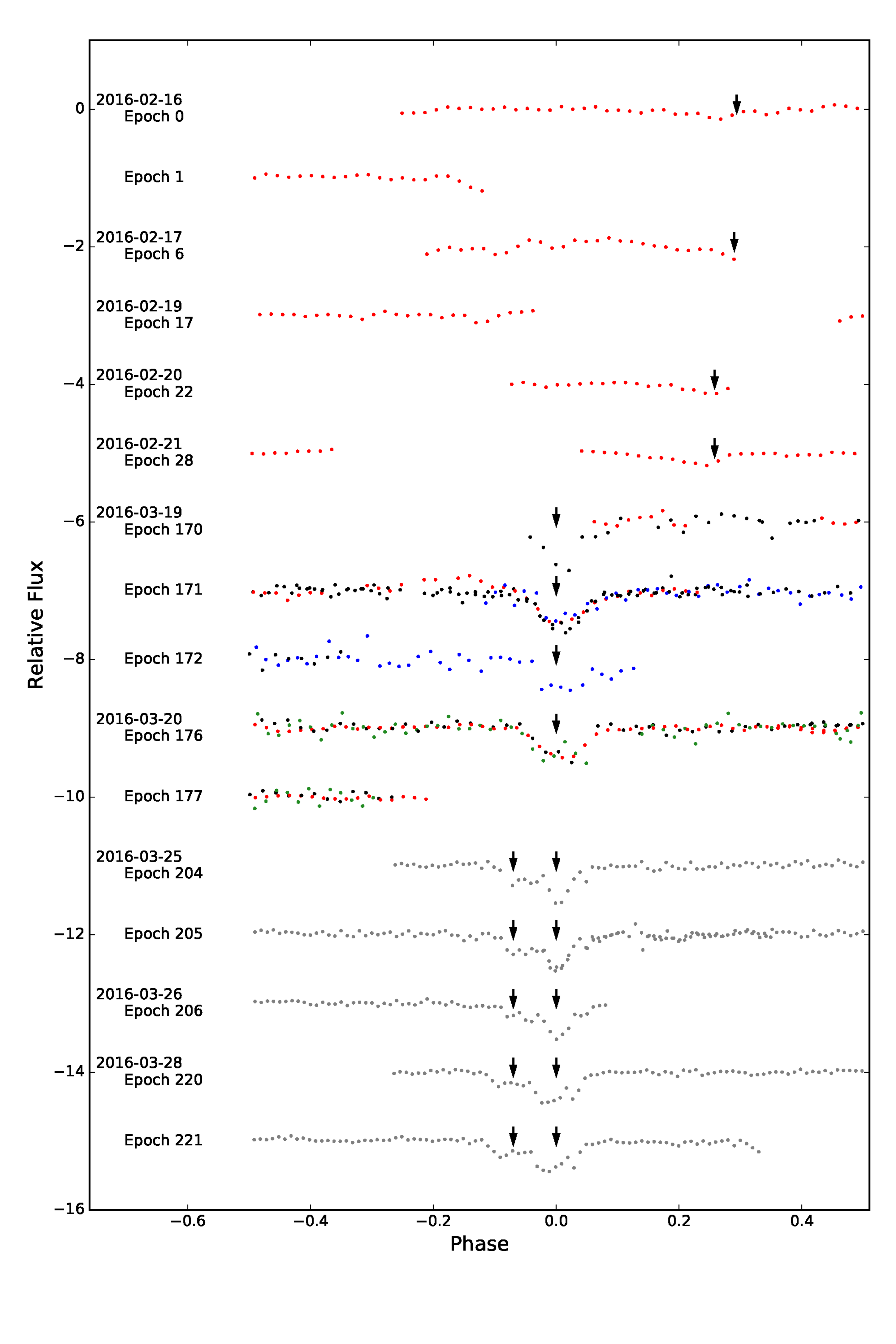}
    \caption{\label{fig:lc_all}Ensemble of light curves presented in this paper, phased to the ephermis presented in Table~\ref{tab:simultaneous_params}. Observations in $J$ band are plotted in red, $R$ band in green, $V$ band in blue, Clear in black, and $g'$ band in grey. The transits modelled in Section~\ref{sec:lightcurve_model} are labelled by the vertical marks.}
\end{figure*}

To provide an overview of the datasets presented in this paper, Figure~\ref{fig:lc_all} plots the ensemble of light curves in chronological order, phased to the transit epoch $T_0$ and period $P$ from Table~\ref{tab:simultaneous_params}.

\section*{Acknowledgements}
\label{sec:acknowledgements}
GZ thanks insightful discussions with Andrew Vanderburg \& Bryce Croll. JPM is supported by a UNSW Vice-Chancellor's Postdoctoral Fellowship. We thank the support staff at the AAO, who made the continued IRIS2 observations possible. This work makes use of observations from the LCOGT network. D. Dragomir acknowledges support provided  by NASA through Hubble Fellowship grant HST-HF2-51372.001-A awarded by the Space Telescope Science Institute, which is operated by the Association of Universities for Research in Astronomy, Inc., for NASA, under contract NAS5-26555.

\bibliographystyle{mn2e}
\bibliography{mybibfile.bib}

\label{lastpage}

\end{document}